\documentclass[journal]{IEEEtran}
\usepackage[ruled, vlined]{algorithm2e}
\usepackage{algpseudocode}
\usepackage{amsmath}
\usepackage{amssymb}
\usepackage{array}
\usepackage{arydshln}
\usepackage{blkarray, bigstrut}
\usepackage{bm}
\usepackage{booktabs}
\usepackage{braket}
\usepackage{cite}
\usepackage{cuted}
\usepackage{enumitem}
\usepackage{graphicx}
\usepackage[mathscr]{euscript}
\usepackage[switch]{lineno}
\usepackage{mathtools}
\usepackage{multirow}
\usepackage{multicol}
\usepackage{framed} 
\usepackage[framed]{ntheorem}
\usepackage{nicefrac}
\usepackage{pgfplots}
\usepackage{qtree}
\usepackage{adjustbox}
\usepackage{rotating}
\pgfplotsset{compat = 1.13}
\usepackage{stfloats}
\usepackage{subfig}
\usepackage{url}
\usepackage{tikz}
\usepackage{tkz-berge}
\usepackage[framemethod=TikZ]{mdframed}
\usetikzlibrary{patterns}
\usetikzlibrary{spy}
\usetikzlibrary{shapes, backgrounds,calc, snakes}
\usetikzlibrary{decorations.pathreplacing}
\usetikzlibrary{matrix}
\usepackage{fancybox}

\tikzstyle{vertex} = [circle, draw, inner sep = 0pt, minimum size = 10pt]

\newframedtheorem{frm-prop}{Property}

\definecolor{bblue}{rgb}{0.12392, 0.0490, 0.9588}
\definecolor{sskyblue}{rgb}{0.1529, 0.5882, 0.9216}
\definecolor{ggreen}{rgb}{0.7098, 0.95, 0.40781}
\definecolor{yyellow}{rgb}{0.9765, 0.9804, 0.0784}

\definecolor{color0}{HTML}{FF0147}
\definecolor{color1}{HTML}{F400DC}
\definecolor{color2}{HTML}{BA0DFF}
\definecolor{color3}{HTML}{5700E8}
\definecolor{color4}{HTML}{0B03FF}
\definecolor{color5}{HTML}{0957F4}
\definecolor{color6}{HTML}{03B3FF}
\definecolor{color7}{HTML}{08E8DA}
\definecolor{color8}{HTML}{07FF8E}
\definecolor{color9}{HTML}{51FF0A}

\definecolor{p1}{rgb}{1, 0.0667, 0}
\definecolor{p2}{rgb}{1, 0.24, 0}
\definecolor{p3}{rgb}{1, 0.349, 0}
\definecolor{p4}{rgb}{1, 0.490, 0}
\definecolor{p5}{rgb}{1, 0.631, 0}
\definecolor{p6}{rgb}{1, 0.792, 0}
\definecolor{p7}{rgb}{1, 0.933, 0}

\begin{document}
	
\title{\huge Decentralized Subchannel Scheduling for C-V2X Mode-4: \\ A Non-Standard Configuration for CAM Retransmissions}

\author{
	\IEEEauthorblockN{Luis F.~Abanto-Leon} \\
	l.f.abanto@ieee.org

}


\maketitle

\begin{abstract}
	In Release 14, 3GPP introduced a novel paradigm known as cellular vehicle--to--everything (C-V2X) \emph{mode-4} to specifically support vehicular communications in scenarios without network coverage. Such a scheme has been devised to operate distributedly harnessing a sensing mechanism whereby vehicles can monitor the received power across subchannels. Based on the measured power intensities, vehicles autonomously select a suitable subchannel over which safety messages are broadcasted. The selected subchannel is utilized by the vehicle for a short period of time before releasing it to select another subchannel. In this work, we propose a scheduling approach based on the aforementioned technology. We assume the existence of a primary sub-band that supports semi-persistently reserved subchannels (as specified by 3GPP). In addition, we also consider that there exist auxiliary sub-bands that convey random re-transmissions in order to improve reception reliability. Different configurations of the proposed approach are explored and their performance are assessed using real vehicular traces. 
\end{abstract}

\begin{IEEEkeywords}
	subchannel scheduling, power sensing, vehicular communications, mode-4, sidelink
\end{IEEEkeywords}

\IEEEpeerreviewmaketitle

\section{Introduction}
The 3rd Generation Partnership Project (3GPP) has included in Release 14, Cellular Vehicle--to--Everything (C-V2X) communications as one of the novel disruptive technologies under the umbrella of 5G. In light of the high dynamicity and unpredictability of the vehicular ecosystem, C-V2X communications is expected to be a dependable technology capable of coping with time-varying vehicular densities while meeting stringent latency and reliability requirements. Guaranteeing such essentialities will prove advantageous for a plethora of use cases, e.g. preventing road traffic accidents \cite{b1}. The type of messages typically exchanged in safety-related applications are periodic cooperative awareness messages (CAMs), which convey information about speed, position, direction, etc. of each vehicle \cite{b2, b9}.

C-V2X outlines two types of scheduling schemes known as \textit{mode-3} and \textit{mode-4}. The former one is a centralized scheme that harnesses cellular infrastructure to suitably distribute the sidelink subchannels among the vehicles in network coverage \cite{b3, b12, b13}. Contrastingly, \textit{mode-4} has been engineered to operate autonomously and distributedly without network support. Thus, vehicles monitor the received power on each subchannel and select an appropriate one for their own usage. Exploiting an strategy of \textit{sensing and selection} \cite{b11, b10, b14, b15}, vehicles attempt to fulfill a two-fold objective: $(ii)$ improve the likelihood of their own messages being received reliably and $(i)$ not affect the conditions of subchannels being utilized by other vehicles. An indisputable advantage of \textit{mode-3} is that subchannels can be more efficiently utilized as eNodeBs can consolidate an humongous knowledge of all vehicles in coverage. Furthermore, due to coherent orchestration between eNodeBs, subchannel assignments free of conflicts can be attained \cite{b5}. Nonetheless, signaling between vehicles and eNodeBs via uplink/downlink may pose a challenge in terms of latency exigencies. Conversely, in \textit{mode-4} due to absence of a central controller, latency owing to signaling and scheduling processing is nonexistent. On the downside, however, limited local knowledge at each vehicle may hinder the selection of a subchannel that is evenly fair for all vehicles in the system. As a consequence of uncoordinated strategies, several vehicles may compete over the same resources thus provoking collisions and leading to severe degradation of the packet reception ratio (PRR). 

In order to diminish the occurrences of conflicts, 3GPP proposed the standardization of a semi-persistent scheduling (SPS) scheme \cite{b4} with the purpose of making the utilization of subchannels more trackable and anticipatable. SPS allows vehicles to comprehend which subchannels are being reserved on a quasi-steady basis. Thus, fostering a more rational selection of subchannels to prevent conflicts. Given the necessity of further boosting reception reliability, we propose a scheduling approach with multiple sub-bands where a primary sub-band mandatorily performs SPS to endow predictability (as defined by 3GPP). In addition, auxiliary sub-bands are defined to effectuate random re-transmissions acting as second-tier supports. To wit, all sub-bands are contained within a 10 MHz intelligent transportation systems (ITS) band. 

The paper is structured as follows. Section II describes the concept of semi-persistent scheduling (SPS) for C-V2X \textit{mode-4}. Section III explains in detail the proposed decentralized scheduling scheme. Section IV is devoted to simulations and discussion. Finally, Section V summarizes our conclusions.

\section{Semi-Persistent Scheduling for C-V2X-Mode 4}
\begin{figure}[!t]
	\centering
	\begin{tikzpicture}[scale = 0.9]
	
	\node[rotate = 90] at (0.2,-0.2) {\tiny $B$ Hz};
	
	\node[rotate = 90] at (0,-0.2) {\tiny sub-band $f$};
	
	\node at (3.1,0) {\dots};
	\node at (3.1,-0.4) {\dots};
	
	\node at (5.3,0) {\dots};
	\node at (5.3,-0.4) {\dots};
	
	\node at (6.7,0) {\dots};
	\node at (6.7,-0.4) {\dots};
	
	\node at (7.7,-0.2) {\dots};
	\node at (8.5,-0.2) {\dots};
	
	\draw[fill=white] (0.4,0) rectangle (0.8,-0.4);
	
	\draw[fill=white] (0.8,0) rectangle (1.2,-0.4);
	
	\draw[pattern=crosshatch, pattern color=color6] (1.2,0) rectangle (1.6,-0.4);
	
	\draw[fill=white] (1.6,0) rectangle (2,-0.4);

	\draw[fill=white] (2,0) rectangle (2.4,-0.4);
	
	\draw[fill=white] (2.4,0) rectangle (2.8,-0.4);
	
	\draw[fill=white] (3.4,0) rectangle (3.8,-0.4);
	
	\draw[fill=white] (3.8,0) rectangle (4.2,-0.4);
	
	\draw[fill=white] (4.2,0) rectangle (4.6,-0.4);

	\draw[pattern=crosshatch, pattern color=color6] (4.6,0) rectangle (5,-0.4);

	\draw[fill=white] (5.6,0) rectangle (6,-0.4);
	
	\draw[fill=white] (6,0) rectangle (6.4,-0.4);

	\draw[fill=white] (7,0) rectangle (7.4,-0.4);

	\draw[very thick] (0.4,0.2) -- (0.4,-0.6);
	\draw[very thick] (3.8,0.2) -- (3.8,-0.6);
	\draw[very thick] (7.4,0.2) -- (7.4,-0.6);
	\draw[very thick] (8.0,0.2) -- (8.0,-0.6);
	
	\node[rotate = 90, align=left, text width = 1cm] at (0.6,0.6) {\tiny $k=1$};
	\node[rotate = 90, align=left, text width = 1cm] at (1,0.6) {\tiny $k=2$};
	\node[rotate = 90, align=left, text width = 1cm] at (1.4,0.6) {\tiny $k=3$};
	\node[rotate = 90, align=left, text width = 1cm] at (1.8,0.6) {\tiny $k=4$};
	\node[rotate = 90, align=left, text width = 1cm] at (2.2,0.6) {\tiny $k=5$};
	\node[rotate = 90, align=left, text width = 1cm] at (2.6,0.6) {\tiny $k=6$};
	\node[rotate = 90, align=left, text width = 1cm] at (3.6,0.6) {\tiny $k=100$};
	\node[rotate = 90, align=left, text width = 1cm] at (4.0,0.6) {\tiny $k=1$};
	\node[rotate = 90, align=left, text width = 1cm] at (4.4,0.6) {\tiny $k=2$};
	\node[rotate = 90, align=left, text width = 1cm] at (4.8,0.6) {\tiny $k=3$};
	\node[rotate = 90, align=left, text width = 1cm] at (7.2,0.6) {\tiny $k=100$};
	
	\node at (6,0.7) {\tiny subchannel};
	\node at (6,0.4) {\scriptsize $s^{(f,k)}$};
	\draw[<-.] (6.2,-0.2) to [out=90, in=180] (6,0.3);

	\draw[decoration={brace, raise=5pt},decorate] (3.78,-0.5) -- node[right=6pt] {} (0.42,-0.5);
	\node at (2.1,-1.05) {\scriptsize $T_w = 100$ ms};
	
	\draw[decoration={brace, raise=5pt},decorate] (7.38,-0.5) -- node[right=6pt] {} (3.82,-0.5);
	\node at (5.6,-1.05) {\tiny message rate-dependent time window};
	
	\draw[decoration={brace, raise=5pt},decorate] (9.08,-0.5) -- node[right=6pt] {} (8.02,-0.5);
	\node at (8.55,-1.05) {\scriptsize $T_w$};
	
	\node at (2.1,-0.55) {\tiny $n=1$};
	\node at (5.6,-0.55) {\tiny $n=2$};
	\node at (8.55,-0.55) {\tiny $n=N_w$};
	
	\draw[decoration={brace, raise=5pt},decorate] (9.1,-1.0) -- node[right=6pt] {} (0.4,-1.0);
	\node at (4.75,-1.5) {\scriptsize $T_{SPS} = N_w T_w$};

	\end{tikzpicture}	
	\caption{SPS operation principle}
	\label{f1}
	\vspace{-0.25cm}
\end{figure}
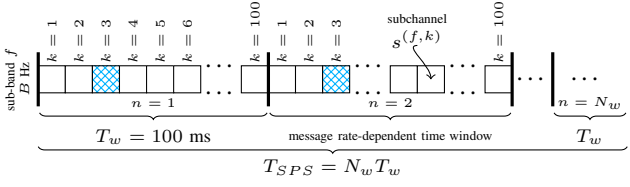

For the purpose of depicting the SPS operation consider Fig. \ref{f1}, which shows a single sub-band $f$. Every sub-band has a bandwidth of $B$ MHz, which has been fragmented into a number of time partitions---hereinafter called subchannels. A subchannel is capable of transporting a safety CAM message and consists of a number of resource blocks (RBs) that are exactly contained within a subframe of 1 ms extent \cite{b4}. Moreover, since a message rate $\Delta_\mathrm{CAM} = 10$ Hz has been assumed \cite{b2}, in each sub-band exist 100 subchannels with periodicity $ T_w = 1 / \Delta_\mathrm{CAM} = 100 $ ms. When a vehicle self-allocates a subchannel in a semi-persistent manner, it will broadcast on such resource periodically during $T_\mathrm{SPS}$ ms, upon whose termination a new reservation will be required. For instance, in Fig. \ref{f1} the subchannel in subframe $k=3$ is being persistently utilized every $T_w$ ms and such reservation pattern remains unchanged during $N_w = T_\mathrm{SPS} / T_w$ consecutive time windows.

\section{Proposed Scheduling Approach}
\begin{figure*}[!b]
	\begin{equation} \label{e1}
	\small
	\varepsilon^{(1,k)}_i =
	\left\{
	\begin{array}{@{}ll@{}}
	\displaystyle \sum_{ \substack{{j=\{ u\mid v_u \in \mathcal{V}^{(k)}\}} \\  u \neq i } } \sum_{p = \{ f \mid s^{(f,k)} \in \mathcal{S}_j\}} I_{p} P_j \frac{G_t \cdot G_r}{\mathcal{X}_{ij} \cdot PL_{ij}}, & \text{if}\ k = \{ m \mid {\mathcal{S}}_i \cap \{ s^{(1,m)},s^{(2,m)},\dots,s^{(F,m)} \} = \emptyset \} \\
	~~~~~~~~~~~~~~~~~~~~~~~~~~\infty, & \text{otherwise}
	\end{array}
	\right.
	\end{equation}
	\vspace{-0.5cm}
\end{figure*}
\begin{figure*}[!t]
	\centering
	\begin{tikzpicture}[scale = 0.85]
	
	\fill[top color=red, bottom color=yellow] (-1.0,0.2) rectangle (-0.5,-2) node[left]{};
	\draw (-1.0,0.2) rectangle (-0.5,-2);
	
	\node at (0.1,0) {\dots};
	\node at (0.1,-0.4) {\dots};
	\node at (0.1,-0.8) {\dots};
	\node at (0.1,-1.4) {\dots};
	\node at (0.1,-1.8) {\dots};
	
	\node at (3.1,0) {\dots};
	\node at (3.1,-0.4) {\dots};
	\node at (3.1,-0.8) {\dots};
	\node at (3.1,-1.4) {\dots};
	\node at (3.1,-1.8) {\dots};

	\node at (6.5,0) {\dots};
	\node at (6.5,-0.4) {\dots};
	\node at (6.5,-0.8) {\dots};
	\node at (6.5,-1.4) {\dots};
	\node at (6.5,-1.8) {\dots};
	
	\node at (7.6,-1) {\dots};
	\node at (8.4,-1) {\dots};
	\node at (9.2,-1) {\dots};
	\node at (10,-1) {\dots};
	
	\node at (13.5,0) {\dots};
	\node at (13.5,-0.4) {\dots};
	\node at (13.5,-0.8) {\dots};
	\node at (13.5,-1.4) {\dots};
	\node at (13.5,-1.8) {\dots};
	
	\node at (16.9,0) {\dots};
	\node at (16.9,-0.4) {\dots};
	\node at (16.9,-0.8) {\dots};
	\node at (16.9,-1.4) {\dots};
	\node at (16.9,-1.8) {\dots};
	
	\node at (18.7,0) {\dots};
	\node at (18.7,-0.4) {\dots};
	\node at (18.7,-0.8) {\dots};
	\node at (18.7,-1.4) {\dots};
	\node at (18.7,-1.8) {\dots};
	
	\node at (0.8,-1.0) {\vdots};
	\node at (1.2,-1.0) {\vdots};
	\node at (1.6,-1.0) {\vdots};
	\node at (2.0,-1.0) {\vdots};
	\node at (2.4,-1.0) {\vdots};
	\node at (2.8,-1.0) {\vdots};
	\node at (3.4,-1.0) {\vdots};
	\node at (4.2,-1.0) {\vdots};
	\node at (4.6,-1.0) {\vdots};
	\node at (5,-1.0) {\vdots};
	\node at (5.4,-1.0) {\vdots};
	\node at (5.8,-1.0) {\vdots};
	\node at (6.2,-1.0) {\vdots};
	\node at (6.8,-1.0) {\vdots};
	\node at (7.2,-1.0) {\vdots};
	\node at (10.8,-1.0) {\vdots};
	\node at (11.2,-1.0) {\vdots};
	\node at (11.6,-1.0) {\vdots};
	\node at (12.0,-1.0) {\vdots};
	\node at (12.4,-1.0) {\vdots};
	\node at (12.8,-1.0) {\vdots};
	\node at (13.2,-1.0) {\vdots};
	\node at (13.8,-1.0) {\vdots};
	\node at (14.6,-1.0) {\vdots};
	\node at (15.0,-1.0) {\vdots};
	\node at (15.4,-1.0) {\vdots};
	\node at (15.8,-1.0) {\vdots};
	\node at (16.2,-1.0) {\vdots};
	\node at (16.6,-1.0) {\vdots};
	\node at (17.2,-1.0) {\vdots};
	\node at (18.0,-1.0) {\vdots};
	\node at (18.4,-1.0) {\vdots};
	
	\node at (3.1,-1.0) {$\ddots$};
	\node at (6.5,-1.0) {$\ddots$};
	\node at (13.5,-1.0) {$\ddots$};
	\node at (16.9,-1.0) {$\ddots$};
	\node at (18.7,-1.0) {$\ddots$};
			
	\draw[fill=white] (0.4,0) rectangle (0.8,-0.4);
	\draw[fill=white] (0.4,-0.4) rectangle (0.8,-0.8);
	\draw[fill=white] (0.4,-1.4) rectangle (0.8,-1.8);
	
	\draw[pattern=horizontal lines, pattern color=black] (0.8,0) rectangle (1.2,-0.4);
	\draw[pattern=crosshatch, pattern color=black] (0.8,-0.4) rectangle (1.2,-0.8);
	\draw[pattern=horizontal lines, pattern color=black] (0.8,-1.4) rectangle (1.2,-1.8);
	
	\draw[fill=black] (1.2,0) rectangle (1.6,-0.4);
	\draw[pattern=horizontal lines, pattern color=black] (1.2,-0.4) rectangle (1.6,-0.8);
	\draw[pattern=horizontal lines, pattern color=black] (1.2,-1.4) rectangle (1.6,-1.8);
	
	\draw[fill=white] (1.6,0) rectangle (2,-0.4);
	\draw[fill=white] (1.6,-0.4) rectangle (2,-0.8);
	\draw[fill=white] (1.6,-1.4) rectangle (2,-1.8);
	
	\draw[fill=white] (2,0) rectangle (2.4,-0.4);
	\draw[fill=white] (2,-0.4) rectangle (2.4,-0.8);
	\draw[fill=white] (2,-1.4) rectangle (2.4,-1.8);
	
	\draw[fill=white] (2.4,0) rectangle (2.8,-0.4);
	\draw[fill=white] (2.4,-0.4) rectangle (2.8,-0.8);
	\draw[fill=white] (2.4,-1.4) rectangle (2.8,-1.8);

	\draw[pattern=horizontal lines, pattern color=black] (3.4,0) rectangle (3.8,-0.4);
	\draw[pattern=horizontal lines, pattern color=black] (3.4,-0.4) rectangle (3.8,-0.8);
	\draw[pattern=crosshatch, pattern color=black] (3.4,-1.4) rectangle (3.8,-1.8);
	
	\draw[fill=white] (3.8,0) rectangle (4.2,-0.4);
	\draw[fill=white] (3.8,-0.4) rectangle (4.2,-0.8);
	\draw[fill=white] (3.8,-1.4) rectangle (4.2,-1.8);
	
	\draw[fill=white] (4.2,0) rectangle (4.6,-0.4);
	\draw[fill=white] (4.2,-0.4) rectangle (4.6,-0.8);
	\draw[fill=white] (4.2,-1.4) rectangle (4.6,-1.8);
	
	\draw[fill=black] (4.6,0) rectangle (5,-0.4);
	\draw[pattern=horizontal lines, pattern color=black] (4.6,-0.4) rectangle (5,-0.8);
	\draw[pattern=horizontal lines, pattern color=black] (4.6,-1.4) rectangle (5,-1.8);
	
	\draw[pattern=horizontal lines, pattern color=black] (5,0) rectangle (5.4,-0.4);
	\draw[pattern=horizontal lines, pattern color=black] (5,-0.4) rectangle (5.4,-0.8);
	\draw[pattern=crosshatch, pattern color=black] (5,-1.4) rectangle (5.4,-1.8);
	
	\draw[pattern=horizontal lines, pattern color=black] (5.4,0) rectangle (5.8,-0.4);
	\draw[pattern=crosshatch, pattern color=black] (5.4,-0.4) rectangle (5.8,-0.8);
	\draw[pattern=horizontal lines, pattern color=black] (5.4,-1.4) rectangle (5.8,-1.8);
	
	\draw[fill=white] (5.8,0) rectangle (6.2,-0.4);
	\draw[fill=white] (5.8,-0.4) rectangle (6.2,-0.8);
	\draw[fill=white] (5.8,-1.4) rectangle (6.2,-1.8);
	
	\draw[fill=white] (6.8,0) rectangle (7.2,-0.4);
	\draw[fill=white] (6.8,-0.4) rectangle (7.2,-0.8);
	\draw[fill=white] (6.8,-1.4) rectangle (7.2,-1.8);
	
	\draw[pattern=horizontal lines, pattern color=black] (10.4,0) rectangle (10.8,-0.4);
	\draw[pattern=horizontal lines, pattern color=black] (10.4,-0.4) rectangle (10.8,-0.8);
	\draw[pattern=crosshatch, pattern color=black] (10.4,-1.4) rectangle (10.8,-1.8);
	
	\draw[fill=p4] (10.8,0) rectangle (11.2,-0.4);
	\draw[fill=white] (10.8,-0.4) rectangle (11.2,-0.8);
	\draw[fill=white] (10.8,-1.4) rectangle (11.2,-1.8);
	
	\draw[fill=black] (11.2,0) rectangle (11.6,-0.4);
	\draw[pattern=horizontal lines, pattern color=black] (11.2,-0.4) rectangle (11.6,-0.8);
	\draw[pattern=horizontal lines, pattern color=black] (11.2,-1.4) rectangle (11.6,-1.8);
	
	\draw[fill=p4] (11.6,0) rectangle (12,-0.4);
	\draw[fill=white] (11.6,-0.4) rectangle (12,-0.8);
	\draw[fill=white] (11.6,-1.4) rectangle (12,-1.8);
	
	\draw[fill=p1] (12,0) rectangle (12.4,-0.4);
	\draw[fill=white] (12,-0.4) rectangle (12.4,-0.8);
	\draw[fill=white] (12,-1.4) rectangle (12.4,-1.8);
	
	\draw[fill=p7] (12.4,0) rectangle (12.8,-0.4);
	\draw[fill=white] (12.4,-0.4) rectangle (12.8,-0.8);
	\draw[fill=white] (12.4,-1.4) rectangle (12.8,-1.8);
	
	\draw[pattern=horizontal lines, pattern color=black] (12.8,0) rectangle (13.2,-0.4);
	\draw[pattern=crosshatch, pattern color=black] (12.8,-0.4) rectangle (13.2,-0.8);
	\draw[pattern=horizontal lines, pattern color=black] (12.8,-1.4) rectangle (13.2,-1.8);
	
	\draw[fill=p4] (13.8,0) rectangle (14.2,-0.4);
	\draw[fill=white] (13.8,-0.4) rectangle (14.2,-0.8);
	\draw[fill=white] (13.8,-1.4) rectangle (14.2,-1.8);
	
	\draw[fill=white] (14.2,0) rectangle (14.6,-0.4);
	\draw[fill=white] (14.2,-0.4) rectangle (14.6,-0.8);
	\draw[fill=white] (14.2,-1.4) rectangle (14.6,-1.8);
	
	\draw[pattern=horizontal lines, pattern color=black] (14.6,0) rectangle (15.0,-0.4);
	\draw[pattern=horizontal lines, pattern color=black] (14.6,-0.4) rectangle (15.0,-0.8);
	\draw[pattern=crosshatch, pattern color=black] (14.6,-1.4) rectangle (15.0,-1.8);
	
	\draw[fill=white] (15.0,0) rectangle (15.4,-0.4);
	\draw[fill=white] (15.0,-0.4) rectangle (15.4,-0.8);
	\draw[fill=white] (15.0,-1.4) rectangle (15.4,-1.8);
	
	\draw[pattern=horizontal lines, pattern color=black] (15.4,0) rectangle (15.8,-0.4);
	\draw[pattern=crosshatch, pattern color=black] (15.4,-0.4) rectangle (15.8,-0.8);
	\draw[pattern=horizontal lines, pattern color=black] (15.4,-1.4) rectangle (15.8,-1.8);
	
	\draw[fill=black] (15.8,0) rectangle (16.2,-0.4);
	\draw[pattern=horizontal lines, pattern color=black] (15.8,-0.4) rectangle (16.2,-0.8);
	\draw[pattern=horizontal lines, pattern color=black] (15.8,-1.4) rectangle (16.2,-1.8);
	
	\draw[fill=white] (16.2,0) rectangle (16.6,-0.4);
	\draw[fill=white] (16.2,-0.4) rectangle (16.6,-0.8);
	\draw[fill=white] (16.2,-1.4) rectangle (16.6,-1.8);
	
	\draw[fill=white] (17.2,0) rectangle (17.6,-0.4);
	\draw[fill=white] (17.2,-0.4) rectangle (17.6,-0.8);
	\draw[fill=white] (17.2,-1.4) rectangle (17.6,-1.8);
	
	\draw[pattern=horizontal lines, pattern color=black] (17.6,0) rectangle (18.0,-0.4);
	\draw[pattern=crosshatch, pattern color=black] (17.6,-0.4) rectangle (18.0,-0.8);
	\draw[pattern=horizontal lines, pattern color=black] (17.6,-1.4) rectangle (18.0,-1.8);
	
	\draw[fill=white] (18.0,0) rectangle (18.4,-0.4);
	\draw[fill=white] (18.0,-0.4) rectangle (18.4,-0.8);
	\draw[fill=white] (18.0,-1.4) rectangle (18.4,-1.8);
	
	\draw[very thick] (0.4,0.2) -- (0.4,-2);
	\draw[very thick] (3.8,0.2) -- (3.8,-2);
	\draw[very thick] (7.2,0.2) -- (7.2,-2);
	\draw[very thick] (8,0.2) -- (8,-2);
	\draw[very thick] (8.8,0.2) -- (8.8,-2);
	\draw[very thick] (9.6,0.2) -- (9.6,-2);
	\draw[very thick] (10.4,0.2) -- (10.4,-2);
	\draw[very thick] (14.2,0.2) -- (14.2,-2);
	\draw[very thick] (17.6,0.2) -- (17.6,-2);
	
	\node at (2.1,-2.1) {\scriptsize $n=1$};
	\node at (5.5,-2.1) {\scriptsize $n=2$};
	\node at (12.3,-2.1) {\scriptsize $n=N_w$};
	\node at (15.9,-2.1) {\scriptsize $n=1$};
	
	\draw[decoration={brace, raise=5pt},decorate] (3.78,-2.9) -- node[right=6pt] {} (0.42,-2.9);
	\node at (2.1,-3.5) {$T_w = 100$ (ms)};
	
	\draw[decoration={brace, raise=5pt},decorate] (7.18,-2.9) -- node[right=6pt] {} (3.82,-2.9);
	\node at (5.5,-3.5) {$T_w = 100$ (ms)};
	
	\draw[decoration={brace, raise=5pt},decorate] (7.98,-2.9) -- node[right=6pt] {} (7.22,-2.9);
	\node at (7.6,-3.5) {$T_w$};
	
	\draw[decoration={brace, raise=5pt},decorate] (8.78,-2.9) -- node[right=6pt] {} (8.02,-2.9);
	\node at (8.4,-3.5) {$T_w$};
	
	\draw[decoration={brace, raise=5pt},decorate] (9.58,-2.9) -- node[right=6pt] {} (8.82,-2.9);
	\node at (9.2,-3.5) {$T_w$};
	
	\draw[decoration={brace, raise=5pt},decorate] (10.38,-2.9) -- node[right=6pt] {} (9.62,-2.9);
	\node at (10,-3.5) {$T_w$};
	
	\draw[decoration={brace, raise=5pt},decorate] (14.18,-2.9) -- node[right=6pt] {} (10.42,-2.9);
	\node at (12.3,-3.5) {$T_w = 100$ (ms)};
	
	\draw[decoration={brace, raise=5pt},decorate] (17.58,-2.9) -- node[right=6pt] {} (14.22,-2.9);
	\node at (15.9,-3.5) {$T_w = 100$ (ms)};
	
	\draw[decoration={brace, raise=5pt},decorate] (14.2,-3.6) -- node[right=6pt] {} (0.4,-3.6);
	\node at (7.3,-4.2) {$T_{SPS}$};
	
	\draw[decoration={brace, raise=5pt},decorate] (0.42,1.0) -- node[right=6pt] {} (7.18,1.0);
	\node at (3.8,1.5) {\scriptsize A SPS subchannel reservation lasts during $T_{SPS}$ ms};

	\draw[decoration={brace, raise=5pt},decorate] (10.42,1) -- node[right=6pt] {} (14.18,1);
	\node at (12.3,2.1) [text width=6cm,align=center] {\scriptsize After monitoring the received power on all};
	\node at (12.3,1.8) [text width=6cm,align=center] {\scriptsize the subchannels of the primary sub-band,};
	\node at (12.3,1.5) [text width=6cm,align=center] {\scriptsize the next SPS subchannel is selected.};
	
	\draw[decoration={brace, raise=5pt},decorate] (14.22,1.0) -- node[right=6pt] {} (18.78,1.0);
	\node at (16.5,1.8) {\scriptsize Newly reserved};
	\node at (16.5,1.5) {\scriptsize SPS subchannel};

	\draw [->] (1.0,-2.18) -- (1.0,-1.9);
	\node at (1.0,-2.34) [text width=2.3cm,align=center] {\scriptsize random};
	\node at (1.0,-2.58) [text width=2.3cm,align=center] {\scriptsize retransmission in};
	\node at (1.0,-2.82) [text width=2.3cm,align=center] {\scriptsize subchannel $s^{(2,2)}$};
	
	\draw [->] (1.4,0.38) -- (1.4,0.1);
	\node at (1.4,1.0) [text width=2.3cm,align=center] {\scriptsize SPS};
	\node at (1.4,0.76) [text width=2.3cm,align=center] {\scriptsize transmission in};
	\node at (1.4,0.52) [text width=2.3cm,align=center] {\scriptsize subchannel $s^{(1,3)}$};
	
	\draw [->] (3.6,-2.18) -- (3.6,-1.9);
	\node at (3.6,-2.34) [text width=2.3cm,align=center] {\scriptsize random};
	\node at (3.6,-2.58) [text width=2.3cm,align=center] {\scriptsize retransmission in};
	\node at (3.6,-2.82) [text width=2.3cm,align=center] {\scriptsize subchannel $s^{(F,100)}$};
	
	\draw [->] (4.8,0.38) -- (4.8,0.1);
	\node at (4.8,1.0) [text width=2.3cm,align=center] {\scriptsize SPS};
	\node at (4.8,0.76) [text width=2.3cm,align=center] {\scriptsize transmission in};
	\node at (4.8,0.52) [text width=2.3cm,align=center] {\scriptsize subchannel $s^{(1,3)}$};
	
	\draw [->] (10.6,-2.18) -- (10.6,-1.9);
	\node at (10.6,-2.34) [text width=2.3cm,align=center] {\scriptsize random};
	\node at (10.6,-2.58) [text width=2.3cm,align=center] {\scriptsize retransmission in};
	\node at (10.6,-2.82) [text width=2.3cm,align=center] {\scriptsize subchannel $s^{(F,1)}$};
	
	\draw [->] (11.4,0.38) -- (11.4,0.1);
	\node at (11.4,1.0) [text width=2.3cm,align=center] {\scriptsize SPS};
	\node at (11.4,0.76) [text width=2.3cm,align=center] {\scriptsize transmission in};
	\node at (11.4,0.52) [text width=2.3cm,align=center] {\scriptsize subchannel $s^{(1,3)}$};
	
	\draw [->] (13,-2.18) -- (13,-1.9);
	\node at (13,-2.34) [text width=2.3cm,align=center] {\scriptsize random};
	\node at (13,-2.58) [text width=2.3cm,align=center] {\scriptsize retransmission in};
	\node at (13,-2.82) [text width=2.3cm,align=center] {\scriptsize subchannel $s^{(2,7)}$};
	
	\draw [->] (16.0,0.38) -- (16.0,0.1);
	\node at (16.0,1.0) [text width=2.3cm,align=center] {\scriptsize SPS};
	\node at (16.0,0.76) [text width=2.3cm,align=center] {\scriptsize transmission in};
	\node at (16.0,0.52) [text width=2.3cm,align=center] {\scriptsize subchannel $s^{(1,5)}$};
	
	\draw[fill=p5] (0,-4.7) rectangle (0.4,-5.1);
	\node at (2.8,-4.8) [text width=4cm,align=left] {\footnotesize Reception with};
	\node at (2.8,-5.1) [text width=4cm,align=left] {\footnotesize power sensing};
	
	\draw[fill=white] (3.4,-4.7) rectangle (3.8,-5.1);
	\node at (6.2,-4.8) [text width=4cm,align=left] {\footnotesize Reception w/o};
	\node at (6.2,-5.1) [text width=4cm,align=left] {\footnotesize power sensing};
	
	\draw[fill=black] (6.8,-4.7) rectangle (7.2,-5.1);
	\node at (9.6,-4.8) [text width=4cm,align=left] {\footnotesize Transmission w/o};
	\node at (9.6,-5.1) [text width=4cm,align=left] {\footnotesize power sensing};
	
	\draw[pattern=crosshatch, pattern color=black] (10.4,-4.7) rectangle (10.8,-5.1);
	\node at (13.2,-4.8) [text width=4cm,align=left] {\footnotesize Retransmission};
	\node at (13.2,-5.1) [text width=4cm,align=left] {\footnotesize w/o power sensing};
	
	\draw[pattern=horizontal lines, pattern color=black] (14.2,-4.7) rectangle (14.6,-5.1);
	\node at (17,-4.8) [text width=4cm,align=left] {\footnotesize Reception / sensing};
	\node at (17,-5.1) [text width=4cm,align=left] {\footnotesize unfeasible due to HD};
	
	\draw (-0.4,-4.5) rectangle (18,-5.3);
	
	\end{tikzpicture}	
	\caption{Proposed scheme with joint SPS scheduling and random retransmissions}
	\label{f2}
\end{figure*}
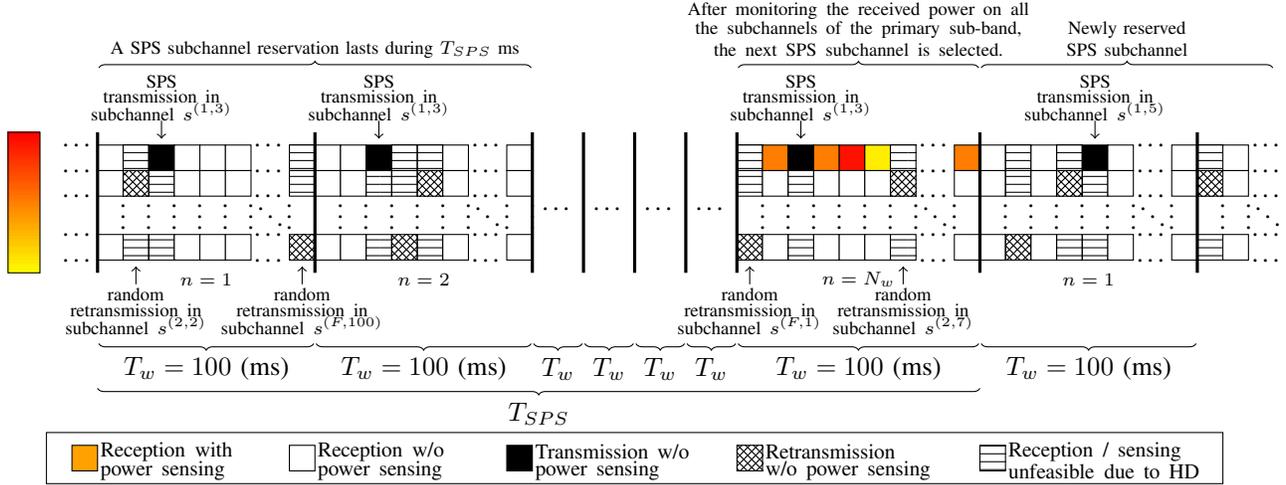

There exist $F$ sub-bands contained within a 10 MHz ITS channel. The primary sub-band reserves subchannels on a semi-persistent basis to add predictability whereas the $F-1$ remaining auxiliary sub-bands execute random retransmissions in order to boost the packet reception reliability. Let $s^{(f,k)}$ denote the subchannel in sub-band $f$ (for $f=1, 2, \dots, F$) and subframe $k$ (for $k=1, 2, \dots, 100$) as depicted in Fig. \ref{f1}. Thus, $\mathcal{S}^{(f)} = \{ s^{(f,1)}, \dots, s^{(f,100)} \}$ represents the complete set of subchannels in sub-band $f$. The proposed scheduling scheme is depicted in Fig. \ref{f2} and consists of the following three phases.
\subsection{Power Sensing}
Because the SPS mechanism has been implemented only in the primary sub-band, power sensing is relevant solely for the subchannels in $\mathcal{S}^{(1)}$. Furthermore, power sensing is performed independently by each vehicle in the last scheduled time window $n = N_w$. Also, since the value of $T_\mathrm{SPS}$ is randomly drawn by each vehicle from a set of predetermined values \cite{b4}, $N_w$ changes at each new scheduling instance on a per-vehicle basis, thus contributing to decorrelating the subchannel selection among vehicles. During any specific time window $n$ (for $n = 1, 2, \dots, N_w$), a vehicle $v_i$ is persistently transmitting a CAM message of size $M_\mathrm{CAM}$ bytes on a determined primary subchannel $\mathcal{S}^\mathrm{prim}_i = \{ s^{(1,\tilde{k}_1)} \}$. Furthermore, in the auxiliary sub-bands, $v_i$ randomly retransmits replicas on the subchannels $\mathcal{S}^\mathrm{aux}_i = \{ s^{(2,\tilde{k}_2)}, s^{(3,\tilde{k}_3)}, \dots, s^{(F,\tilde{k}_F)} \}$, for $ \tilde{k}_f \in \{1, 2, \dots, 100\} $. Due to half-duplex PHY assumption, vehicle $v_i$ will be able to receive signals and sense power intensities only when not transmitting. For instance, if vehicle $v_i$ is transmitting a replica on subchannel $s^{(\tilde{f},\tilde{k})}$ for $\tilde{f} \neq 1$, the received power on the corresponding primary subchannel $s^{(1,\tilde{k})}$ will be unknown. As a result, there will be at most $F$ primary subchannels where power sensing will not be feasible. Based on the foregoing, the primary SPS subchannel for the forthcoming scheduling period will be chosen from among $100-F$ candidates (at most). The power $\varepsilon^{(1,k)}_i$ perceived by vehicle $v_i$ at each primary subchannel $s^{(1,k)}$ is computed by (\ref{e1}), where $\mathcal{V}^{(k)}$ represents the set of all the vehicles $v_j$ broadcasting on primary/auxiliary subchannels of subframe $k$. On the other hand, ${\mathcal{S}}_i = {\mathcal{S}}^\mathrm{prim}_i \cup {\mathcal{S}}^\mathrm{aux}_i$ denotes the set of alloted subchannels to vehicle $v_i$ in the last time window $n = N_w$. The transmit power of vehicle $v_j$ is represented by $P_j = P_T$, which is assumed to be the same for all units. The antenna gains of the transmitter and receiver are $G_t$ and $G_r$, respectively. The parameter $\mathcal{X}_{ij}$ is a log-normal random variable with standard deviation $\mathcal{X}_{\sigma}$ representing shadow fading experienced by the link between vehicles $v_i$ and $v_j$. In addition, $PL_{ij} = \max \{PL_{ij}^{\text{free-space}}, PL_{ij}^{B_1}\} $ depicts the path loss between $v_i$ and $v_j$, where the first and second terms denote power attenuation based on free-space and $\mathrm{WINNER+ UMi~(B\textsubscript{1})}$ \cite{b8} models, respectively. $I_{p}$ is a weighting factor to depict the power level contributed by subchannels ${\mathcal{S}}_j$ onto subchannel $s^{(1,k)}$. Notice that for every vehicle $v_j \in \mathcal{V}^{(k)}$, ${\mathcal{S}}_j \cap \{ \cup_{p=1}^F s^{(p,k)} \}\neq \emptyset$ holds. Thus, if $\mathcal{S}^\mathrm{prim}_j \neq \emptyset$ then $s^{(p=1,k)}$ is being utilized by $v_j$ and therefore $I_{p=1} = 1$. On the other hand, if $ \mathcal{S}^\mathrm{aux}_j \neq \emptyset$, then $v_j$ is broadcasting on at least one auxiliary subchannel, in which case $I_{p \neq 1} \leq 1$. This former case represents the influence of in-band emissions (IBE), i.e. power leaking from auxiliary subchannels $s^{(p \neq 1,k)}$ to the primary subchannel $s^{(1,k)}$. Regardless of the utilized sub-band, when vehicle $v_i$ is transmitting on subframe $k$, the sensed power on the corresponding primary subchannel $s^{(1,k)}$ is unknown. Therefore, $\varepsilon^{(1,k)}_i$ is set to $\infty$ in order to preclude its reselection in the following phase.

\subsection{Subchannel Selection for Semi-Persistent Broadcasting}
Once the received power intensities have been measured in $\mathcal{S}^{(1)}$, each vehicle $v_i$ will rank the primary subchannels in ascending order. Let such a sorted set be denoted by $\widetilde{\mathcal{S}}_{i}^{100} = \{ \tilde{s}_{i,1}, \tilde{s}_{i,2}, \dots, \tilde{s}_{i,100} \}$, where $\tilde{s}_{i,l} \leq \tilde{s}_{i,l+1}$. In a more general manner, let $\widetilde{\mathcal{S}}^K_i = \{ \tilde{s}_{i,1}, \tilde{s}_{i,2}, \dots, \tilde{s}_{i,K} \}$ denote the first $K$ primary subchannels of the sorted set $\widetilde{\mathcal{S}}_{i}^{100}$. Thus, the subchannel selection process consists on each vehicle $v_i$ randomly choosing one subchannel from $\widetilde{\mathcal{S}}^K_i$ for its own SPS transmission. Notice that when $K = 1$, it yields a greedy scheme, where the subchannel with lowest power is selected. When $K = 100$, the selection becomes purely random. In the former case, several vehicles experiencing similar subchannel conditions may unknowingly select the same resources and broadcast concurrently. Thus, leading to an increased amount of collisions. In the latter case, the rate of messages colliding might diminish but---on the downside---subchannels with high interference may be selected. In both cases, extreme values of $K$ will lead to PRR degradation. In this work, $K$ remains uniform for all the vehicles. Nevertheless, $K$ plays an important role as it influences both the amount of colliding messages and the optimality of the reserved subchannels. 
\subsection{CAM Retransmissions}
Each vehicle $v_i$ will randomly broadcast a retransmission in a subchannel $s^{(f,\tilde{k}_f)} \in \mathcal{S}^\mathrm{aux}_i$ at each of the $F-1$ auxiliary sub-bands with the purpose of improving reception reliability.

\section{Simulations}
In this section, the influence of parameter $K$ on the selectivity of candidate SPS primary subchannels is firstly evaluated. Subsequently, the impact of retransmissions on PRR performance is assessed. For the realized experiments, a high vehicle density region of the \emph{TAPAS Cologne database} \cite{b6} was chosen, where an average number of 3000 vehicles over 40 seconds was observed. In addition, the relevant parameters for all the experiments are shown in Table \ref{t1}.
\begin{center}
	\begin{table}[!h]
		\centering
		\scriptsize
		\caption {Simulation parameters}
		\label{t1}
		\begin{tabular}{lccc}
			\toprule
			\multicolumn{1}{c}{\textbf{Description}} & \multicolumn{1}{c}{\textbf{Symbol}} & \multicolumn{1}{c}{\textbf{Value}} & \multicolumn{1}{c}{\textbf{Units}}\\
			\midrule
			Number of RBs per subchannel & - & 32 & - \\
			Number of sub-bands & $F$ & {1,2,3} & - \\
			Number of subchannels per sub-band & - & 100 & -\\
			Subchannel selectivity index & $K$ & 1-100 & -\\
			CAM message rate & $\Delta_\mathrm{CAM}$ & 10 & Hz \\
			CAM size & $M_\mathrm{CAM}$ & 200 & bytes\\
			MCS & - & 6 & - \\
			Transmit power & $P_T$ & 23 & dBm \\
			Effective coded throughput (24 CRC bits) & $\rho$ & 0.916 & bps$\slash$Hz \\
			Throughput loss coefficient \cite{b7} & $\lambda$ & 0.6 & - \\
			SINR threshold & $\gamma_{T}$ & 2.75 & dB \\
			Distance between Tx and Rx & $D_x$ & 50-300 & m \\
			Scheduling period \cite{b4} & $T_\mathrm{SPS}$ & 0.5-1.5 & s \\
			Sensitivity threshold (per subchannel) & - & -103.4 & dBm \\
			Antenna gain & $G_t, G_r$ & 3 & dB \\
			Shadowing standard deviation & $\mathcal{X}_{\sigma}$ & 7 & dB \\
			Shadowing correlation distance & - & 10 & m \\
			\bottomrule
		\end{tabular}
	\end{table}
	\vspace{-0.75cm}
\end{center}

\subsection{Scenario I: Impact of K on PRR}
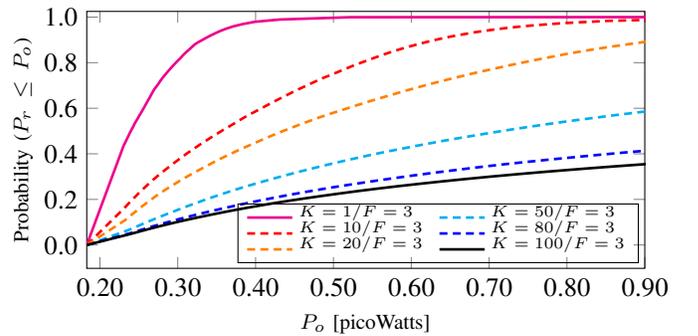
\begin{figure}[!t]
	\centering
	\begin{tikzpicture}
	\begin{axis}[
	xmin = 0.1831,
	xmax = 0.9,
	ymax = 1.025,
	width = 9.0cm,
	height = 5.0cm,
	xlabel={$P_o$ [picoWatts]},
	x label style={align=center, font=\footnotesize,},
	xtick = {0.20, 0.30, 0.40, 0.50, 0.60, 0.70, 0.80, 0.90},
	xticklabels = {0.20, 0.30, 0.40, 0.50, 0.60, 0.70, 0.80, 0.90},
	ylabel = {Probability ($P_r \leq P_o$)},
	y label style={at={(-0.08,0.5)}, text width = 3.5cm, align=center, font=\footnotesize,},
	ytick = {0.0, 0.2, 0.4, 0.6, 0.8, 1.0},
	yticklabels = {0.0, 0.2, 0.4, 0.6, 0.8, 1.0},
	legend columns = 2,
	legend style={at={(0.27,0.02)},anchor=south west, font=\fontsize{6}{5}\selectfont, text width=1.78cm,text height=0.02cm,text depth=.ex, fill = none, align = left},
	]
	
	\addplot[color=magenta, mark options = {fill = yyellow}, line width = 1pt, style = solid] table {ReceivedEnergyK1.txt}; \addlegendentry{$K = 1 / F = 3$}
	
	\addplot[color=cyan, mark options = {fill = yyellow}, line width = 1pt, style = densely dashed] table {ReceivedEnergyK50.txt}; \addlegendentry{$K = 50 / F = 3$}
	
	\addplot[color=red, mark options = {fill = yyellow}, line width = 1pt, style = densely dashed] table {ReceivedEnergyK10.txt}; \addlegendentry{$K = 10 / F = 3$}
		
	\addplot[color=blue, mark options = {fill = yyellow}, line width = 1pt, style = densely dashed] table {ReceivedEnergyK80.txt}; \addlegendentry{$K = 80 / F = 3$}
	
	\addplot[color=orange, mark options = {fill = yyellow}, line width = 1pt, style = densely dashed] table {ReceivedEnergyK20.txt}; \addlegendentry{$K = 20 / F = 3$}

	\addplot[color=black, mark options = {fill = yyellow}, line width = 1pt, style = solid] table {ReceivedEnergyK100.txt}; \addlegendentry{$K = 100 / F = 3$}
	
	\end{axis}
	\end{tikzpicture}
	\caption{CDF of received power on primary subchannels}
	\label{f3}
	\vspace{-0.25cm}
\end{figure}

Fig. \ref{f3} portrays how the cumulative distribution function (CDF) of the received power $P_r$ on the primary subchannels changes with $K$. We have assumed that $F=3$ and therefore both co-channel interference (CCI) and in-band emissions were considered. Although Fig. \ref{f3} may suggest that choosing a small value of $K$ is more appropriate from a single vehicle perspective, it constitutes an adverse strategy that may impinge on the system performance. Intuitively, when $K$ is relatively small the selection tends to become greedy and therefore packet collisions are more prone to occur. Conversely, when $K$ is large, the randomization improves the decorrelation of subchannel allocation among vehicles---which reduces the amount of collision occurrences. Nevertheless, due to stochastic selection and low selectivity, subchannels with high interference might be reserved provoking severe PRR degradation. Several simulations were conducted in order to find a suitable value of $K$ capable of providing a balanced trade-off between the amount of generated collisions and the optimality of candidate subchannels. Thus, it was discovered that values between 20 and 35 are capable of providing superior PRR performance. 

\subsection{Scenario II: Impact of retransmissions on PRR}
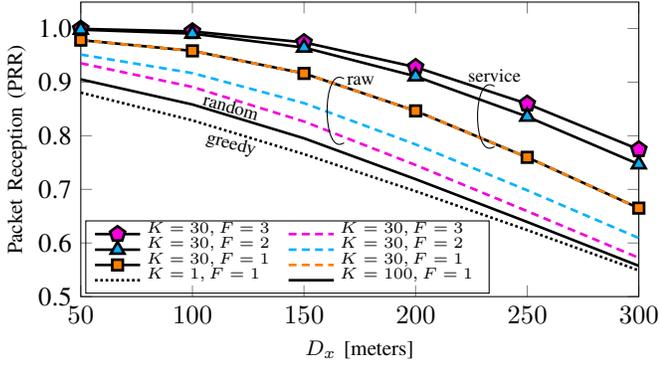
\begin{figure}[!]
	\centering
	\begin{tikzpicture}
	\begin{axis}[
	xmin = 50,
	xmax = 300,
	ymin = 0.5,
	width = 9.0cm,
	height = 5.5cm,
	xlabel={$D_x$ [meters]},
	x label style={align=center, font=\footnotesize,},
	ylabel = {Packet Reception (PRR)},
	y label style={at={(-0.08,0.5)}, text width = 3.5cm, align=center, font=\footnotesize,},
	ytick = {0.5, 0.6, 0.7, 0.8, 0.9, 1.0},
	yticklabels = {0.5, 0.6, 0.7, 0.8, 0.9, 1.0},
	legend columns = 2,
	legend style={at={(0.008,0.01)},anchor=south west, font=\fontsize{6}{7}\selectfont, text width=1.8cm,text height=0.01cm,text depth=.ex, fill = none, align = left},
	]
	
	\addplot[color = black, mark = pentagon*, mark options = {scale = 1.5, fill = color1, solid}, line width = 1pt] coordinates 
	{
		(50, 0.999320)
		(100, 0.994576)
		(150, 0.974207)
		(200, 0.927993)
		(250, 0.859636)
		(300, 0.774514)
	}; \addlegendentry{$K=30, F=3$}
	
	\addplot[color=color1, mark options = {fill = yyellow}, line width = 1pt, style = densely dashed] coordinates 
	{
		(50, 0.935234)
		(100, 0.891040)
		(150, 0.826591)
		(200, 0.745414)
		(250, 0.659402)
		(300, 0.572354)	
	}; \addlegendentry{$K=30, F=3$}

	\addplot[color=black, mark = triangle*, mark options = {scale = 1.5, fill = color6}, line width = 1pt] coordinates 
	{
		(50, 0.997785)
		(100, 0.990486)
		(150, 0.964508)
		(200, 0.910867)
		(250, 0.835695)
		(300, 0.746828)
	}; \addlegendentry{$K=30, F=2$}

	\addplot[color=color6, mark options = {fill = yyellow}, line width = 1pt, style = densely dashed] coordinates 
	{
		(50, 0.951396)
		(100, 0.916871)
		(150, 0.860828)
		(200, 0.783914)
		(250, 0.698227)
		(300, 0.609292)
	}; \addlegendentry{$K=30, F=2$}

	\addplot[color=black, mark = square*, mark options = {fill = orange, solid}, line width = 1pt] coordinates
	{
		(50, 0.978345)
		(100, 0.958347)
		(150, 0.916263)
		(200, 0.846390)
		(250, 0.759755)
		(300, 0.665047)
	}; \addlegendentry{$K=30, F=1$}

	\addplot[color=orange, mark options = {fill = yyellow}, line width = 1pt, style = densely dashed] coordinates 
	{
		(50, 0.978345)
		(100, 0.958347)
		(150, 0.916263)
		(200, 0.846390)
		(250, 0.759755)
		(300, 0.665047)
	}; \addlegendentry{$K=30, F=1$}
	
	\addplot[color=black, mark options = {fill = yyellow}, line width = 1pt, style = densely dotted] coordinates 
	{
		(50, 0.880457)
		(100, 0.828685)
		(150, 0.765703)
		(200, 0.696585)
		(250, 0.624121)
		(300, 0.548834)	
	}; \addlegendentry{$K=1, F=1$}

	\addplot[color=black, mark options = {fill = yyellow}, line width = 1pt, style = solid] coordinates 
	{
		(50, 0.905139)
		(100, 0.858340)
		(150, 0.795272)
		(200, 0.719052)
		(250, 0.639250)
		(300, 0.557368)	
	}; \addlegendentry{$K=100, F=1$}

	\end{axis}
	
	\draw (5.5,2.0) .. controls (5.2,1.8) and (5.2,3) .. (5.5,2.8);
	\node at (5.5,2.95) {\scriptsize service};
	
	\draw (3.5,2.05) .. controls (3.2,1.85) and (3.2,3.1) .. (3.5,2.9);
	\node at (3.725,2.88) {\scriptsize raw};
	
	\node[rotate = -15] at (2.0,2.03) {\scriptsize greedy};
	\node[rotate = -16] at (2.0,2.53) {\scriptsize random};
	
	\end{tikzpicture}
	\caption{PRR for different distances between vehicles}
	\label{f4}
	\vspace{-0.2cm}
\end{figure}

Considering $K=30$, in this scenario the proposed scheme with a single primary sub-band ($F = 1$) is contrasted against the cases where one ($F = 2$) and two ($F = 3$) auxiliary sub-bands are utilized. Fig. \ref{f4} illustrates these three cases for different awareness distances $D_x$ and compares them on the basis of two PRR variants; namely $PRR_\mathrm{raw}$ and $PRR_\mathrm{service}$. The former variant computes the PRR as if every CAM message (i.e. nominal transmissions and retransmissions) were independent. When $F=\{2,3\}$, message replicas in auxiliary sub-bands take place blindly and therefore the chosen subchannels might exhibit high CCI; thus impinging on $PRR_\mathrm{raw}$ performance. Conversely, when $F=1$ every transmitted message will likely occur on pre-selected high-quality primary subchannels. As a result, the performance of $PRR_\mathrm{raw}$ will be higher when $F = 1$ than in the cases where $F \neq 1$. On the other hand, $PRR_\mathrm{service}$ does take into account that CAM messages in the auxiliary sub-bands are replicas. Therefore, if at least one of the $F$ transmitted messages is recovered correctly the PRR can be leveraged. As a consequence, $PRR_\mathrm{service}$ exhibits enhanced performance over $PRR_\mathrm{raw}$. 

Notice that as more auxiliary sub-bands are utilized, the performance of $PRR_\mathrm{service}$ is boosted. The advantage of broadcasting an additional re-transmission (i.e. $F=2$) is noticeable since a gain of 7\% was obtained over the simpler setting for $D_x = 300$. Furthermore, the additional gain when using $F=3$ is 2\% if compared to $F=2$. The reason for this modest improvement is the presence of in-band emissions that may leak unwanted power onto adjacent subchannels. This phenomenon was partly mitigated for the case $F=2$ by only utilizing the first and third sub-bands. For the sake of comparison, Fig. \ref{f4} also contrasts the following three cases: ($i$) $K = 30 \slash F = 1$, ($ii$) $K = 100 \slash F = 1$ (random SPS) and ($iii$) $K = 1 \slash F = 1$ (greedy SPS). Based on the foregoing, it is manifest that power sensing followed by subchannel selection has the potential to improve the PRR performance when $K$ is chosen adequately. Furthermore, it was observed that under the presence of mild shadowing (e.g. $\mathcal{X}_{\sigma} = 3$), the performance of random SPS is more apparent than greedy SPS as a gain of 8\% was observed for $D_x = 300$. 

\begin{equation} \label{e2}
\small
\gamma^{(f,k)}_{ij} = \frac{P_j g_{ij}}{ \displaystyle \sum_{ \substack{{l=\{ u\mid v_u \in \mathcal{V}^{(k)} \} } \\ { l \neq j, l \neq i} } } \sum_{p = \{ f \mid s^{(f,k)} \in \mathcal{S}_l\}} I_{f,p} \cdot P_l g_{il} + \sigma^2 },
\end{equation}

The PRR is computed for each distance $D_x$ checking whether every pair of vehicles $v_i$ and $v_j$ is within the awareness distance or not. If affirmative, the signal--to--interference--plus--noise ratio (SINR) $\gamma^{(f,k)}_{ij}$ experienced by $v_i$ upon reception of a message transmitted by $v_j$ on subchannel $s^{(f,k)}$ is computed via (\ref{e2}), where $g_{ij} = \frac{G_t \cdot G_r}{\mathcal{X}_{ij} \cdot PL_{ij}}$ and $\sigma^2$ represents the noise power experienced by the receiver at $v_i$. For $F = 3$, the weighting factor $I_{f,p}$ is defined as the element in position ${\mid p-f+1 \mid}$ of the vector $\mathbf{I} = [1~10^{-3}~10^{-4}]$. Thus, $\gamma^{(f,k)}_{ij}$ is compared against a threshold $\gamma_{T} = 10 \cdot \log_{10}(2^{\rho/\lambda}-1)$ \cite{b7}, which is derived from the parameters in Table \ref{t1}. It is assumed that a message can be correctly decoded if its SINR is larger than the mentioned threshold.

Considering the $PRR_\mathrm{service}$ metric and $D_x=300$, it was observed that when $F=1$, the amount of undecodable packets due to excessive CCI was 26.9\%. Approximately 6.9\% of the packets was lost owing to propagation conditions. When $F=2$ and $F=3$, the effective amount of packets not recovered due two severe CCI and IBE was 17.1\% and 12.2\%, respectively. This reveals that the probability of a message failing successively due to CCI decreases as the number of retransmissions is increased. And since replicas take place in the auxiliary sub-bands, only IBE increases but CCI in the primary sub-band is not worsened. A small component of lost packets was due to the half-duplex limitation. For instance, the amount of missed packets for $F=1$ was 0.4\% whereas for $F=2$ and $F=3$ the numbers decrease to 0.0138\% and 0.0006\%, respectively. 

\section{Conclusions}
In this work we proposed a semi-persistent scheduling scheme endowed with multiple sub-bands for improving reception reliability. The described scheme was tested in a scenario with high vehicle density and we could conclude that sensing-based scheduling can provide superior performance over other more simplistic strategies such as random selection. However, the former can only be guaranteed when the subchannel selectivity index is properly devised. Due to implementation of the SPS mechanism, it is feasible to predict and infer future usage patterns from neighboring vehicles and thus reduce the amount of collisions. Finally, we could corroborate that blind random retransmissions on the auxiliary sub-bands have the potential to boost PRR performance as the amount of effective collisions and half-duplex impairments tend to reduce.

\end{document}